\documentclass{article}
\usepackage{amsmath,amssymb,booktabs}
\usepackage{graphicx}
\usepackage[version=3]{mhchem}
\usepackage[colorlinks=true,linkcolor=black,citecolor=black,urlcolor=black]{hyperref}
\input{preamble}
\begin{document}

\title{Uncertainty for calculating transport on Titan: a probabilistic description
        of bimolecular diffusion parameters}

\author{S.~Plessis\\
\PECOS\\
\texttt{splessis@ices.utexas.edu}
\and
D.~McDougall\\
\PECOS
\and
K.~Mandt\\
\SwRI\\
\texttt{kmandt@swri.edu}
\and
T.~Greathouse\\
\SwRI
\and
A.~Luspay-Kuti\\
\SwRI
}

\maketitle

\begin{abstract}
Bimolecular diffusion coefficients
are important parameters used by atmospheric
models to calculate altitude profiles of minor constituents in an
atmosphere.  Unfortunately, laboratory measurements of these coefficients were
never conducted at temperature conditions relevant to the atmosphere of Titan.
Here we conduct a detailed uncertainty analysis of the bimolecular diffusion
coefficient parameters as applied to Titan's upper atmosphere to provide a
better understanding of the impact of uncertainty for this parameter on models.
Because temperature and pressure conditions are
much lower than the laboratory conditions in which bimolecular diffusion
parameters were measured, we apply a Bayesian framework, a problem-agnostic
framework, to determine parameter estimates and associated uncertainties.  We
solve the Bayesian calibration problem using the open-source \QUESO\ library
which also performs a propagation of uncertainties in the calibrated parameters
to temperature and pressure conditions observed in Titan's upper atmosphere.
Our results show that, after propagating uncertainty through the Massman model,
the uncertainty in molecular diffusion is highly
correlated to temperature and we observe no noticeable correlation with
pressure.  We propagate the calibrated molecular diffusion estimate and
associated uncertainty to obtain an estimate with uncertainty due to
bimolecular diffusion for the methane molar fraction as a function of altitude.
Results show that the uncertainty in methane abundance due to molecular
diffusion is in general small compared to eddy diffusion and the chemical
kinetics description.  However, methane abundance is most sensitive to
uncertainty in molecular diffusion above $1200$~km where the errors are
nontrivial and could have important implications for scientific research based on
diffusion models in this altitude range.

\end{abstract}

%%% Keywords
% \begin{keyword}
% Titan \sep diffusion \sep planetary atmosphere \sep uncertainties
% \end{keyword}

\section{Introduction}
Titan atmospheric models contain two major modules: a chemical kinetics module
and a transport module.  While there have been numerous efforts to quantify
uncertainty in the chemical kinetics module
\cite{Hebrard06,Hebrard07,Carrasco07a,Carrasco07b,Carrasco2008,Dobrijevic2008a,
Hebrard2009,Plessis2010,Peng2010,Gans2013}, the transport module is not as
thoroughly evaluated.  Furthermore, the transport module consists of two
components: an eddy diffusion component and a molecular diffusion component.
\cite{Carrasco07c} assessed the question of uncertainty in chemical transport,
but only focussed on the eddy diffusion profile effects, since it is the main
contributor to uncertainty in the lower atmosphere, \cite[Sec.~2.2.3]{WilsonPhD}
determined by a Monte Carlo approach a best fit of eddy diffusion profile.
However, molecular
diffusion plays a critical role in determining altitude profiles in the upper
atmosphere and has yet to be investigated cognizant of uncertainty and
sensitivity.  Parameters for molecular diffusion---molecular binary diffusion
coefficients for each pair of molecules---have not been revisited since their
initial measurements made in 1973 \cite{Wakeham1973}.

\begin{figure}
\centering
\includegraphics{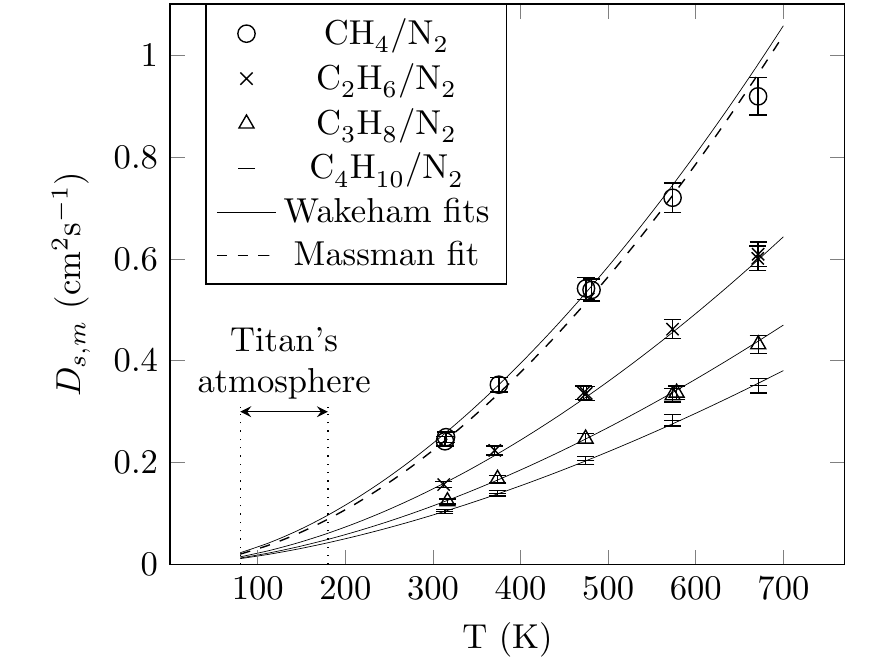}
\caption{\label{WakeMassPlot}Measurements and fits.  The uncertainty in the
laboratory measurements is 4\%. Titan's atmospheric conditions are highlighted
to emphasize the extrapolation required to apply these values to modeling
Titan's atmosphere.}
\end{figure}

\cite{Wilson04} reviewed binary diffusion, describing three different sources
for the binary molecular diffusion coefficients, all of which are the result of
a nonlinear least-squares fit to experimental measurements.  These measurements
were performed under temperatures ranging from $300$~K to $700$~K at 1~atm (see
Fig.~\ref{WakeMassPlot}).  In order to model bimolecular
diffusion in Titan's atmosphere, measurements must be extrapolated to conditions
relevant to Titan---approximately $150$~K and $10^{-6}$~atm.  Here we evaluate the
potential error involved in propagating to temperature and pressure relevant to
Titan's atmosphere by also propagating its uncertainty through this extrapolation.

In this work we apply a probabilistic calibration framework based on Bayes' theorem.
The application of this Bayesian framework solves the problem of
estimating, with uncertainty, unknown molecular diffusion parameters in a model
for Titan's atmosphere.  We review a brief history of the evolution of
bimolecular diffusion models in section 2, and in section 3 we set up the
Bayesian framework in which we estimate parameters for bimolecular diffusion.
In section 4 we detail the measurements and measurement error of bimolecular
diffusion available in the scientific literature and we couple these
measurements with a Bayesian calibration procedure to estimate parameters for
bimolecular diffusion.  This estimate is then propagated through a bimolecular
diffusion model to obtain an estimate of bimolecular diffusion of the
\ce{N2}/\ce{CH4} pair in physical conditions consistent with those typically
found on Titan's atmosphere.  Using this, we then estimate the mixture
diffusion coefficient of \ce{N2}.  We execute this entire process while taking
into account all available experimental and a priori uncertainty.  Section 5
concludes and summarises the paper.

\section{Molecular diffusion models}
Molecular diffusion in Titan's atmosphere is
modeled using the Wilke equation \cite{Wilke1950,Wilson04},
or the mixture-average rule, from the bimolecular
diffusion coefficient:
\begin{equation}
D_s = \frac{ n_\text{total} - n_s}
           {\sum_{j_m \neq s} \frac{n_{j_m}}{\Dbin[s,j_m]}},
\label{mol_diff}
\end{equation}
where $n_\text{total}$ and $n_s$ are the total molar concentration and molar
concentration of species $s$, respectively, and \Dbin[s,j_m]\ is the bimolecular
diffusion coefficient between species $s$ and species $j_m$.  The $j_m$
notation is used to take into account a possible approximation of the
atmosphere's composition where only the dominant species are considered, thus
defining a ``medium'' in which the molecules are diffusing.  In Titan's
atmosphere, this medium is typically composed of \ce{N2}, \ce{CH4} and
sometimes \ce{H2}.

\cite{Haye2005} derived a modified version of the
mixture-averaged rule applicable for gas mixtures with more than two
components.  The purpose of this derivation was to address the rapidly changing
composition of Titan's upper atmosphere above the homopause, where binary
diffusion is dominant.  It required a rederivation of the transport
term due to molecular diffusion, adapted to minor species diffusing through
a bulk gas,
\begin{equation}
\omega^{D}_s = -D_s \left(  \frac{1}{n_s}\frac{\partial n_s}{\partial z}
                          + \frac{1}{H_s}
                          + (1 + \alpha_T)\frac{1}{T}\frac{\partial T}{\partial z}
                  \right),
\label{D:orig}
\end{equation}
to a modified version given by
\begin{equation}
\omega^{D}_s = - \tilde{D_s}
                 \left(  \frac{1}{n_s}\frac{\partial n_s}{\partial z}
                          + \frac{1}{H_s}
                          + (1 + \frac{n_\text{total} - n_s}{n_\text{total}}\alpha_T)\frac{1}{T}\frac{\partial T}{\partial z}
                 \right),
\label{D:modif}
\end{equation}
where
  $H_s$ is the scale height of species $s$,
  $\alpha_T$ is the thermal coefficient,
  $n_\text{total}$ is the total density of the atmosphere and
\begin{equation}
\tilde{D}_s = \frac{D_s}{1 - \frac{n_s}{n_\text{total}} (1 - \frac{m_{s}}{m_{\neq s}}) },
\end{equation}
where
  $m_{s}$ is the molecular mass of species $s$ and
  $m_{\neq s}$ is the mean molecular
    mass of the atmosphere without species $s$. This expression of molecular
diffusion is suited for both minor and major species, with Eq.~\ref{D:modif}
converging towards Eq.~\ref{D:orig} for the minor species.
The detailed analysis is given in \cite[Section~3.3.1]{Haye2005}.

Three different models have been proposed in the literature
for bimolecular diffusion between any pair of molecules,
denoted as the pair of molecules $s$ and $m$. Note that a
binary diffusion coefficient is specific to a pair of molecules
and is agnostic to which species is diffusing and which species is the medium.
First, \cite{Wakeham1973} proposed a model depending only on
temperature (Eq.~\ref{Eq:Wakeham}),
\begin{equation}
\Dbin(T) = A T^s,
\label{Eq:Wakeham}
\end{equation}
where $T$ is the temperature and $A$ and $s$ are the model parameters.
Note that the temperature should be a reduced unitless temperature:
it is more appropriate to write $\frac{T}{\Tref}$ instead of $T$
with $\Tref = 1$~K\@.
The second model, by \cite{Wilson04}, introduced a pressure dependence,
\begin{equation}
\Dbin(T,P) =  \frac{A}{n_\text{tot}}T^\beta,
\label{Eq:Wilson}
\end{equation}
where $n_\text{tot}$ is the total molar concentration, $A$ and $\beta$
are the parameters. The pressure dependence
is given by the mixture equation of state which, in the case of Titan's
atmosphere, is the ideal gas law, $n_\text{tot} = P/\mathrm{R}T$.  Note
that, just as above, it is more appropriate to write this with respect to a
reference temperature.
Finally, \cite{Massman1998} also proposed a model with explicit
temperature and pressure dependence, but included the bimolecular diffusion,
\Dzo, at $0^\circ$C and $1$~atm (\Tz\ and \Pz\ respectively) as a reference
value,
\begin{equation}
\Dbin(T,P) = \Dzo \frac{\Pz}{P}\left(\frac{T}{\Tz}\right)^{\beta}.
\label{Eq:Massman}
\end{equation}
with \Dzo\ and $\beta$ being the parameters of the model.

Modeling Titan's atmosphere requires a pressure extrapolation,
it is therefore more appropriate to consider either the model of
\cite{Massman1998} or
\cite{Wilson04}. Those models are equivalent through the state
equation. From a modeler's point-of-view, the model of
\cite{Massman1998} has
the advantage to extrapolate a reference value of the diffusion
coefficient, which holds therefore the potential to be informed
independently.

The rest of this work uses the model of \cite{Massman1998}.

%
%In the case of unknown experimental data, we
%extrapolate the self-diffusion value
%of the medium species. Bimolecular
%diffusion between medium species $m$ and uncharacterized
%species $s$ is given by:
%\begin{equation}
%\begin{split}
%\beta_{s,m}   & = \beta_{m,m} \\
%D_{s,m}(0,1)  & = \left\{\begin{array}{ll}
%                           D_{m,m}(0,1)\sqrt{\frac{\frac{\Mm{s}}{\Mm{m}} + 1}{2}} & \text{if } \Mm{s}  <   \Mm{m} \\[5pt]
%                           D_{m,m}(0,1)\sqrt{\frac{\Mm{s}}{\Mm{m}}}               & \text{if } \Mm{s} \geq \Mm{m} \\
%                         \end{array}
%                  \right.
%\end{split}
%\end{equation}

\section{Bayesian framework}
In many scientific applications, some quantities need to be determined
from experimental observations that were subject
to errors during the experiment.  Since the quantity is an extrapolation of
measurements, an exact value of this quantity cannot be known with
certainty.  It is therefore important to provide a measure of the associated
uncertainty.  We do this for the bimolecular diffusion model described above in
a Bayesian setting.  The Bayesian setting is advantageous because it provides a
\emph{distribution} of a quantity of interest, and this is achieved in two
steps.  The first step is calibration, where the model is fit to the data to
obtain distributions for model parameters.  The second step propagates these
distributions to a quantity of interest. The Bayesian approach is
not a statistical tool, but rather a method for describing an inverse problem.
It differs from the simple Monte Carlo technique commonly used in planetary
science applications. In the simple Monte Carlo technique one would use a Monte
Carlo model as a tool to cast random experimental measurements from the
measured mean values and errors, and then apply a Massman model fitting to each
set of randomly determined measurements in order to obtain a probability
distribution of \Dzo\ or $\beta$. In the Bayesian approach, we use a Markov
chain Monte Carlo (MCMC) algorithm (a statistical tool) to solve a Bayesian
inverse problem. This approach is very similar to the simple Monte Carlo
technique except that the MCMC does not cast random observations, as described
above, but \Dzo\ and $\beta$. These proposed values are then compared to the
observations and the Metropolis-Hastings acceptance probability decides whether
or not the proposed $(\Dzo, \beta)$ sample should be kept. The samples of \Dzo\
and $\beta$ at the end of the procedure are realizations from the probability
distribution $p(\Dzo, \beta | y)$, which is desired output for our study.

For the calibration step, we need to describe the model to calibrate and the
data to calibrate against.  We will calibrate the Massman model for bimolecular
diffusion~(Eq.~\ref{Eq:Massman}) against direct noisy observations of \Dbin\
at various temperatures and pressures,
\begin{equation}
  y_{jk} = \Dbin(T_j, P_k) + \eta_{jk},
  \quad \eta_{jk} \sim \mathcal{N}(0, \sigma_{jk}),
  \label{eq:likelihood}
\end{equation}
where $j = 1, \ldots, J$, and $k = 1, \ldots, K$.  For the sake of simplicity,
we write the observations as a column vector, $y = (y_{11}, \ldots,
y_{JK})^{\top}$.  Here $\mathcal{N}(0, \sigma^2)$ is notation for a Gaussian
distribution with mean zero and variance $\sigma^2$.  The $\sigma_{jk}$ will
appear later in the expression for the likelihood (see
Section~\ref{sec:likelihood}).

The calibration step proceeds by finding the `best' model parameters, \Dzo\
and $\beta$, given the observations $y$.  That is, we seek the joint
probability distribution $p(\beta, \Dzo | y)$.  Applying Bayes's
theorem yields,
\begin{equation}
  \underbrace{p(\beta, \Dzo | y)}_{\text{posterior}} \propto
  \underbrace{p(y | \beta, \Dzo)}_{\text{likelihood}}
  \underbrace{p(\beta,\Dzo)}_{\text{prior}}.
  \label{diffusion:Bayes}
\end{equation}
The prior distribution encodes a previously held state of knowledge, or an
external expert opinion, about what the parameters $\beta$ and \Dzo\ should
look like.  It is therefore given.  The likelihood distribution is also known;
given $\beta$ and \Dzo, one inserts them into the Massman model and evaluates
Eq.~\ref{eq:likelihood}.  Therefore, since the right-hand side of
Eq.~\ref{diffusion:Bayes} is known, the left-hand side is also known.
The posterior distribution is the solution to the Bayesian
calibration problem and there are many numerical methods to understand its
properties.  We choose to approximate the posterior distribution by statistical
samples (see Section \ref{sec:mcmc}).  Although we have applied the Bayesian
calibration framework to a specific model, in general the framework is
problem-agnostic and can be used in any scientific domain for parameter
estimation and uncertainty quantification.

The second step is to propagate the posterior distribution above, which we
choose to approximate by samples, to the quantity of interest.  For our
purposes, the quantity of interest is the bimolecular diffusion
$\Dbin(\tilde{T}, \tilde{P})$ at some desired temperature
$\tilde{T}$ and pressure $\tilde{P}$\@.  This is very easily done by evaluating
\begin{equation}
  \Dzo_n \frac{\Pz}{\tilde{P}} \left( \frac{\tilde{T}}{\Tz} \right)^{\beta_n}
\end{equation}
for each sample $n = 1, \ldots, N$\@.  This will yield another sequence of
samples of bimolecular diffusions at the desired temperature and pressure.  An
estimate of the bimolecular diffusion at temperature $\tilde{T}$ and pressure
$\tilde{P}$ can be obtained by computing the sample mean.  The associated
uncertainty of this estimate is obtained by computing the sample variance.

\subsection{Choice of prior}
The choice of a prior distribution is a very important part of the Bayesian
framework, as it has a direct influence on the posterior's description
(see Eq.~\ref{diffusion:Bayes}).
In general, this choice can be very difficult.  It
is a statement about knowledge of the solution of the problem, which is
unknown.  However there are situations in which one can provide a prior
distribution without knowing the solution exactly, but knowing some
\emph{property} of the solution.  For example, an external observer can never
know the exact speed of a car but it is assuredly non-negative and less than
the speed of light.  Although contrived, this example illustrates that common
sense can be harnessed when expert guidance is absent.

% Domain expert
When a domain expert is present, one may attempt to acquire high-quality
information which can be incorporated into a prior.  Of course, the aim here is
to do this as accurately as possible.  This is still an active area of
research for which no gold standard exists \cite{Jenkinson05}.  The process of
obtaining expert opinion for a particular problem is called
\emph{elicitation}.

% MEP
When one truly possesses no problem insight, the prior typically relies on the
maximum entropy principle (MEP) \cite{Ryu1993,Maxent97}.  This principle
states that, given a set of constraints, the distribution that encodes the
least amount of information is one which maximizes the Shannon entropy
\cite{Shannon1948}.  For instance, if a mean and standard deviation of a
parameter are given as constraints, the distribution that maximizes the Shannon
entropy is a Gaussian.  This method provides a distribution that attempts to
maximize ignorance about the solution of the problem.  Of course, when a domain
expert is available one can still utilise the maximum entropy principle by
setting the constraints through expert elicitation.

The parameter \Dzo\ is a diffusion coefficient and must be positive whilst
$\beta$ is an exponent and therefore can take any real value.  With no
knowledge other than bounds, the probability distribution that maximises
ignorance in the Shannon entropy sense is a uniform
distribution.  Therefore, a priori, we choose $\Dzo \sim U[0, \infty]$ and
$\beta \sim U[-\infty, \infty]$ so the prior distribution is a joint uniform on
the $(\Dzo, \beta)$ pair.

\subsection{Likelihood}
\label{sec:likelihood}
In Eq.~\ref{diffusion:Bayes} we must evaluate the likelihood of observing the
data $y$ given values of the parameters \Dzo\ and $\beta$.  Since the
experimental data was taken at a fixed set of temperatures $\mathbf{T} \in
\mathbb{R}^J$ and pressures $\mathbf{P} \in \mathbb{R}^K$, we define the
function $\mathcal{G}$ to be,
\begin{align*}
  \mathcal{G}(\Dzo, \beta) &= \Dbin(\mathbf{T}, \mathbf{P}) \\
  &= \Dzo \frac{\Pz}{\mathbf{P}}\left(\frac{\mathbf{T}}{\Tz}\right)^{\beta},
\end{align*}
where division, multiplication and exponentiation are all done component-wise
on the vectors $\mathbf{T}$ and $\mathbf{P}$.  We then derive the likelihood
by recalling from Eq.~\ref{eq:likelihood} that the difference of the model
output and the observation is a Gaussian random variable.  Therefore, the
likelihood distribution function is given by a Gaussian PDF,
\begin{equation}
  p(y | \beta, \Dzo)
  = \frac{1}{Z}
  \exp \left(-\frac{1}{2}\left(\mathcal{G}(\beta, \Dzo) - y\right)^{\top}
                           \Sigma^{-1}
                         \left(\mathcal{G}(\beta, \Dzo) - y\right) \right).
\label{bayes:likelihood}
\end{equation}
The coefficient $Z$ is a constant of proportionality and need not be computed.
The matrix $\Sigma$ is called the error covariance matrix and is equal to
$\diag{\sigma_{11}, \ldots, \sigma_{JK}}$, where $\sigma_{jk}$ is as in
Eq.~\ref{eq:likelihood}.
In general the expression of the likelihood function is determined by the
knowledge available about the probability distribution of the experimental
errors.  Since most experimental results consist of an average of many
measurements, and averages typically follow a Gaussian distribution, it is
commonplace to express the likelihood as a Gaussian \cite{GUM}.

\subsection{Markov chain Monte Carlo}
\label{sec:mcmc}
The Bayesian framework described above is nothing more than a general problem
statement expressing the distribution of model parameters given noisy
observations in terms of a prior and a likelihood.  To solve problems posed in
this framework, numerical methods are used to understand the shape of the
posterior, its mean, its variance, or to compute probabilities.  We choose to
use a method called Markov chain Monte Carlo to understand the posterior
distribution on $\beta$ and \Dzo.

Given a general probability distribution function $p(x)$, Markov chain Monte
Carlo (MCMC) methods produce a sequence of samples that, when plotted via a
histogram, approximate $p(x)$.  In the limit of infinite samples, the
approximate distribution converges to $p$.  Most MCMC
methods rely on the Metropolis-Hasting algorithm \cite{Metropolis1953}.

In our case $p$ will be the posterior distribution function in the Bayesian
framework (Eq.~\ref{diffusion:Bayes}) and the MCMC will result in
$N$ samples, $\{ \beta_n, \Dzo_n
\}_{n=1}^{N}$, from the posterior distribution.  One can compute the mean and
variance of these samples to obtain an estimate with associated uncertainty.
There are many variants of this algorithm each with their own benefits and
drawbacks \cite{Cotter2012,Roberts2001,Beskos2009,Roberts1997,Roberts1998}.
Initialising the Markov chain is also a challenging task, since a bad initial
choice can be detrimental to sample quality \cite{Kathryn2014,Plummer2006}.
Typically, the initial condition is a guess provided by the user and therefore
may lie in a low probability region with respect to the posterior distribution.
In what follows, we maximise the posterior distribution to find a
MAP (maximum a posteriori) point before starting the sampler.  This
ensures the Markov chain starts in stationarity.

Samples in this work were generated by the delayed-rejection
adaptive-Metropolis MCMC algorithm (DRAM) \cite{Haario2006} implemented in the
\QUESO\ library \cite{Prudencio2012}.  \QUESO\ is a C++ library for
quantifying uncertainty in Bayesian calibration problems.  It supports
large-scale models that use many processors, and utilises multi-core
architectures to provide high-quality samples from probability distributions
using Markov chain Monte Carlo.  \QUESO\ is free and open source software,
available at \UrlQUESO.

\section{Results}
\subsection{Data available}
The data available for the diffusivity of the
\CNcouple\ couple are the measurements and fits of
\cite{Wakeham1973} and the fits of \cite{Massman1998}.

It is clear that the prior distribution influences the resulting posterior
distribution (see Eq.~\ref{diffusion:Bayes}).
We chose a minimally informed prior distribution, obtained
by solving the maximum entropy principle.
We assert that \Dzo\ is real and non-negative
and $\beta$ is real.  This provides
uninformative uniform priors $\Dzo \sim U[0, \infty]$ and
$\beta \sim U[-\infty, \infty]$.

\cite{Wakeham1973} performed their measurements
with pressure $P_\text{(Wa)} = 1~\text{atm}$. Using the
ideal gas assumption, we have the following conversion equations,
\begin{equation}
\begin{split}
\beta_\text{(Ma)} & = \beta_{\text{(Wa)}}, \\
\Dzo              & = A_{\text{(Wa)}} \frac{P_\text{(Wa)}}{\Pz} \left(\frac{\Tz}{\Tref}\right)^{\beta_{\text{(Wa)}}}.
\end{split}
\label{conv}
\end{equation}
where $A_{\text{(Wa)}}$ and $\beta_{\text{(Wa)}}$ are the \cite{Wakeham1973} model
parameters, $\beta_{\text{(Ma)}}$ and \Dzo\ the \cite{Massman1998} model parameters.
\begin{table}
\centering
\begin{tabular}{ccc}\toprule
$T$ (K)  &  \DCN\ (cm$^2$\,s$^{-1}$) & Relative error (\%)\\\midrule
313.7    & 0.242                                      & 4 \\
314.9    & 0.250                                      & 4 \\
375.2    & 0.353                                      & 4 \\
474.7    & 0.542                                      & 4 \\
481.0    & 0.539                                      & 4 \\
573.5    & 0.720                                      & 4 \\
671.1    & 0.919                                      & 4 \\
\bottomrule
\end{tabular}

\caption{\label{data_Wakeham}Data from \cite{Wakeham1973} used for the calibration.}
\end{table}

\subsection{Modeling conditions}
\begin{figure}
\centering
\includegraphics[width=0.49\linewidth]{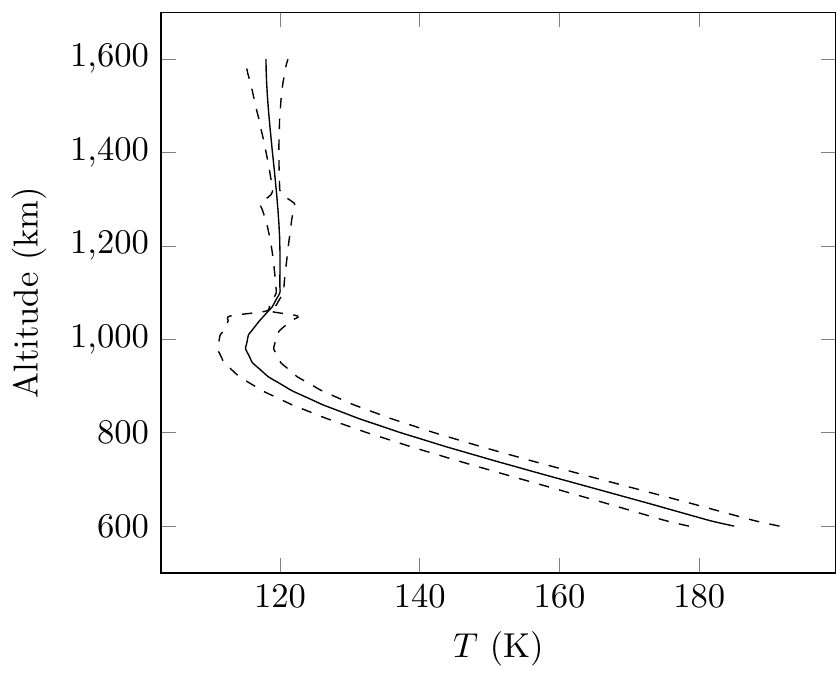}\hfill
\includegraphics[width=0.49\linewidth]{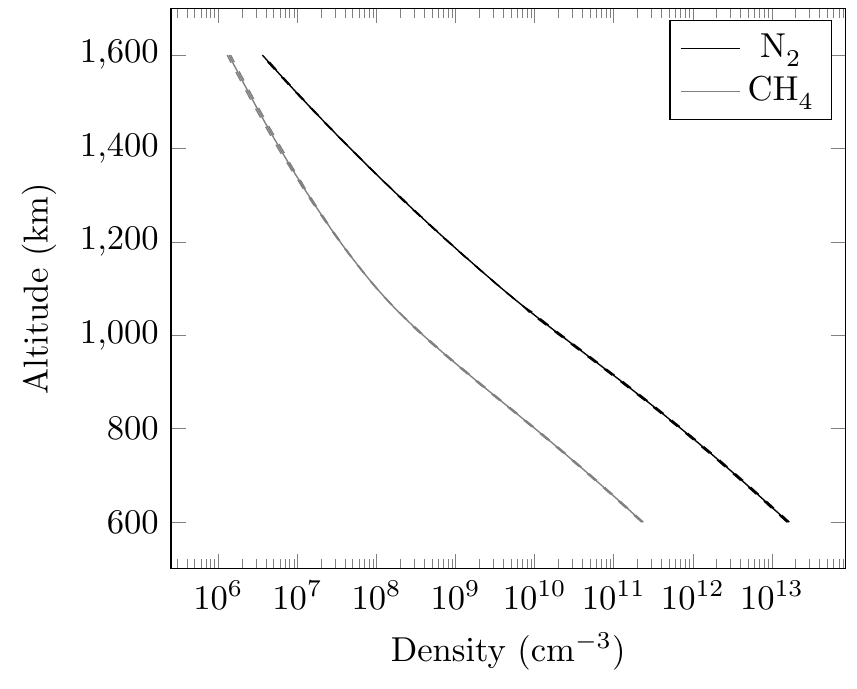}
\caption{Estimated temperature profile of the T40 flyby (left panel) and
         densities of \ce{N2} and \ce{CH4} for this flyby (right panel).
        The uncertainties are given at a $1-\sigma$ level.}
\label{cond_T40}
\end{figure}

The temperature profile and medium species density conditions under which our
calibration and uncertainty propagation have been performed were chosen to
match those derived from data obtained during Cassini's 40th flyby of Titan,
commonly referred to as T40. The temperature profile, shown in Fig.~\ref{cond_T40}, 
is based on fits of diffusion models to density measurements made by instruments 
on the Cassini orbiter and the Huygens lander as described in% \cite{Mandt2012,Mandt2012b}.

\subsection{Calibration and convergence of calibrated quantities}
\begin{figure}
\centering
\includegraphics[width=0.49\textwidth]{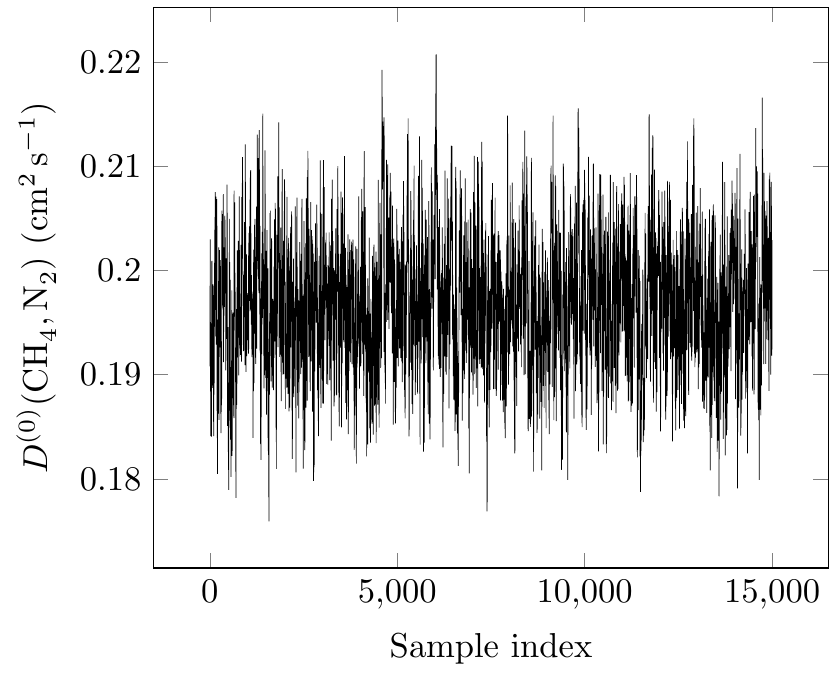}\hfill
\includegraphics[width=0.49\textwidth]{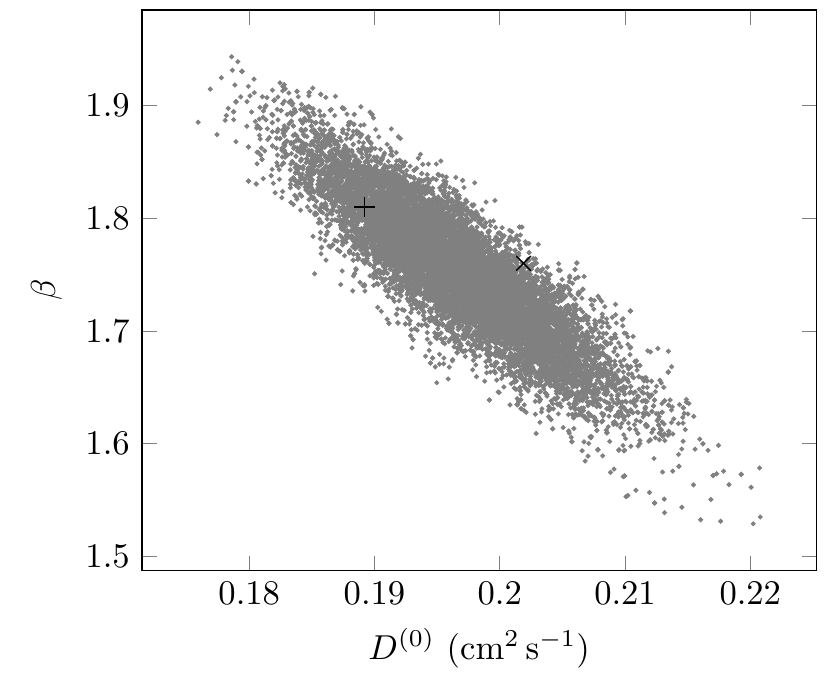}
\caption{\label{post_sample}Left: first $10^4$ output sample draws for \Dzo\ parameters;
right: \CNcouple\ bimolecular diffusivity calibration. The starting
point is determined by a maximization of the likelihood function.
The rank correlation is $-0.85$. Black cross: \cite{Wakeham1973}
fitted values, black plus: \cite{Massman1998} fitted values.}
\end{figure}

\begin{figure}
\centering
\includegraphics[width=\textwidth]{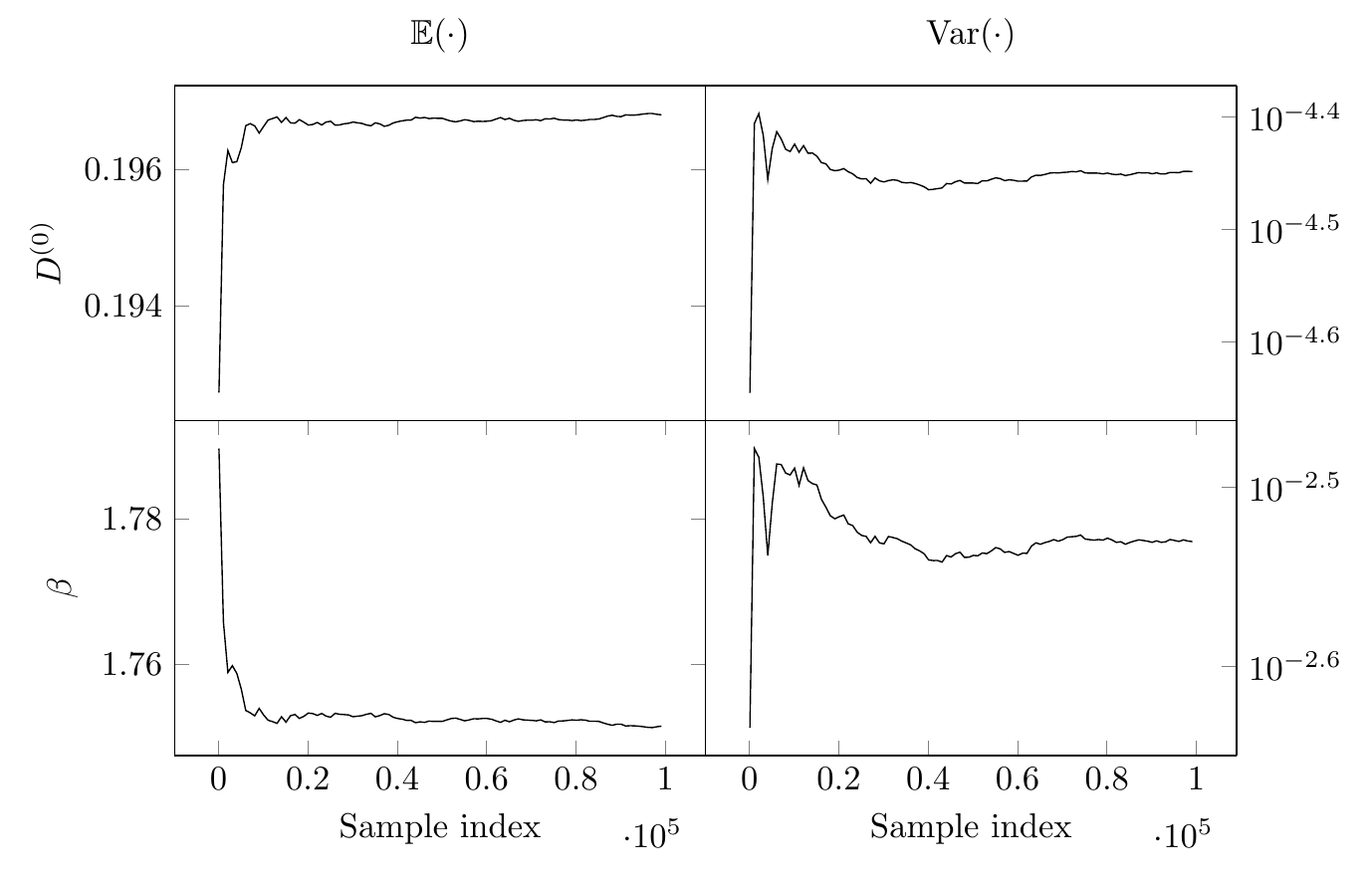}
\caption{\label{stats_output}Means and variances on the generated samples.}
\end{figure}

Fig.~\ref{post_sample} shows $10^4$ samples from the posterior distribution
of \Dzo.
To assess how adequately a finite number of samples represent the posterior
distribution, one typically investigates the convergence of the
moments of the sampled variables with respect to sampler iteration.  The
moments we are interested in are the mean and variance of $\Dzo|_y$ and
$\beta|_y$ (Fig.~\ref{stats_output}).

Fig.~\ref{stats_output} shows the mean ($\mathbb{E}$)
and variance ($\text{Var}$) have converged. As expected, the mean converges faster than the
variance.

\begin{figure}
\centering
\includegraphics[width=0.49\textwidth]{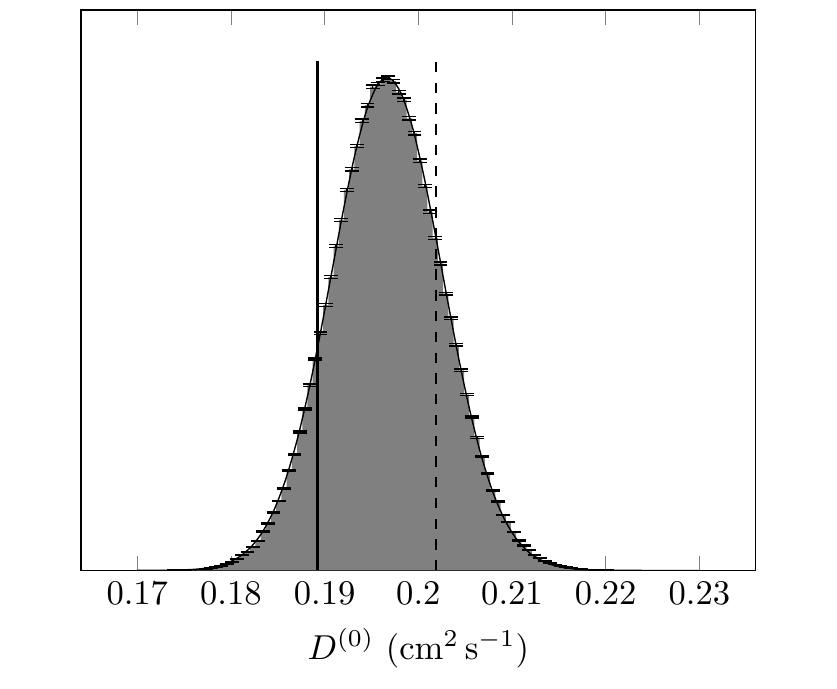}\hfill
\includegraphics[width=0.49\textwidth]{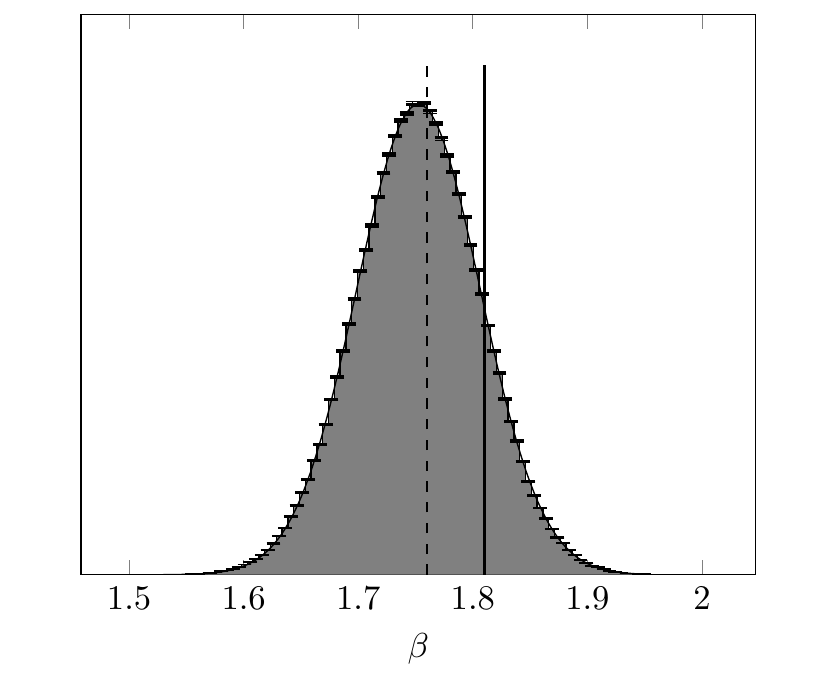}
\caption{\label{posteriors}Posterior histograms. Vertical dashed line: \cite{Wakeham1973} fitted
values, vertical solid line: \cite{Massman1998} fitted values.}
\end{figure}

The marginal distributions are approximated by histograms and
Gaussian fits in Fig.~\ref{posteriors}. The
distributions contain the values of the previous
fits of \cite{Wakeham1973} and \cite{Massman1998}
(Tab.~\ref{Fit}).
\begin{table}
\centering
\renewcommand{\arraystretch}{1.3}
\begin{tabular}{c@{\hspace{10pt}}ccc}\toprule
Model       & \Dzo     (cm$^2$\,s$^{-1}$) & $\beta$         \\\midrule
Wakeham fit & $0.2019$                    & $1.76$          \\
Massman fit & $0.1892$                    & $1.81$          \\
This work   & $0.197 \pm 0.006$           & $1.75 \pm 0.05$ \\\bottomrule
\end{tabular}

\caption{\label{Fit}Comparison of the parameters' value. The corresponding
Massman parameters are given for the Wakeham value using the conversion
equation~\ref{conv}}
\end{table}

\subsection{Propagation and convergence of propagated quantities}

\begin{figure}
\begin{minipage}{0.48\linewidth}
\includegraphics[width=\textwidth]{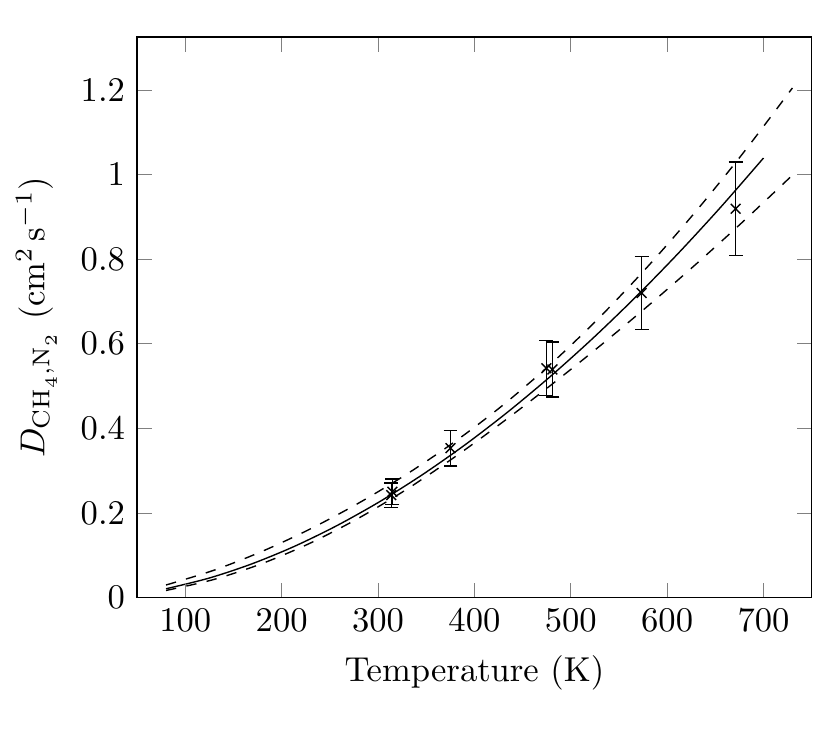}
\caption{\label{prog_D12}Estimated bimolecular diffusivity of
\CNcouple\ at 1~atm. The dashed lines show the
$3-\sigma$ interval.}
\end{minipage}\hfill
\begin{minipage}{0.48\linewidth}
\centering
\includegraphics[width=\textwidth]{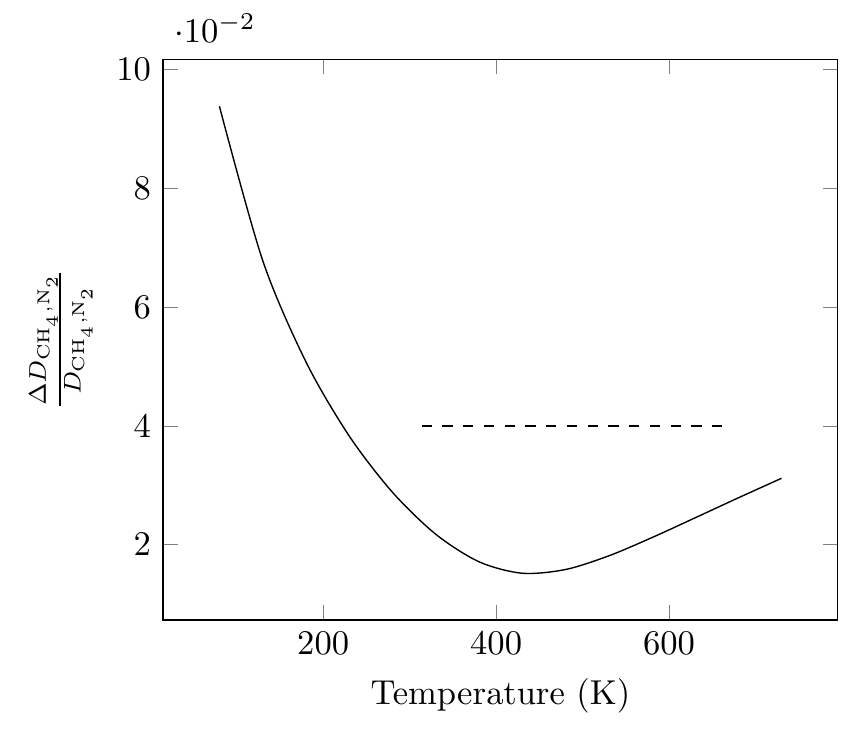}
\caption{\label{prog-unc_D12}Relative uncertainty of the bimolecular diffusivity
of the couple \CNcouple\ at 1~atm. Solid: estimated, dashed: measurements.}
\end{minipage}
\end{figure}

\begin{figure}
\begin{minipage}{0.48\linewidth}
\includegraphics[width=\textwidth]{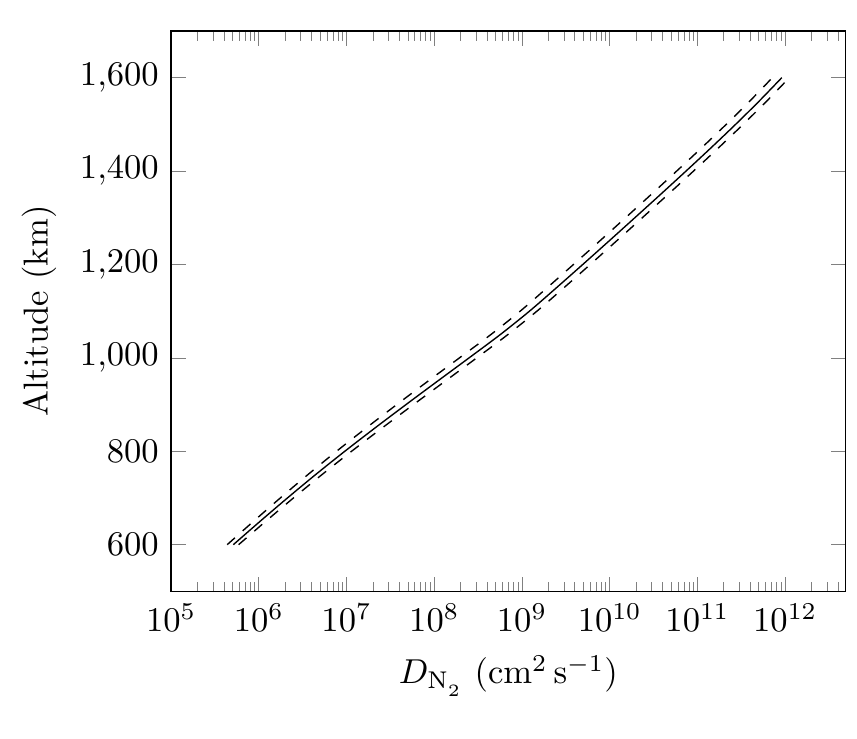}
\caption{\label{prog_D}Wilke diffusivity of \ce{N2}, the
$3-\sigma$ interval is shown with the dash lines.}
\end{minipage}\hfill
\begin{minipage}{0.48\linewidth}
\centering
\includegraphics[width=\textwidth]{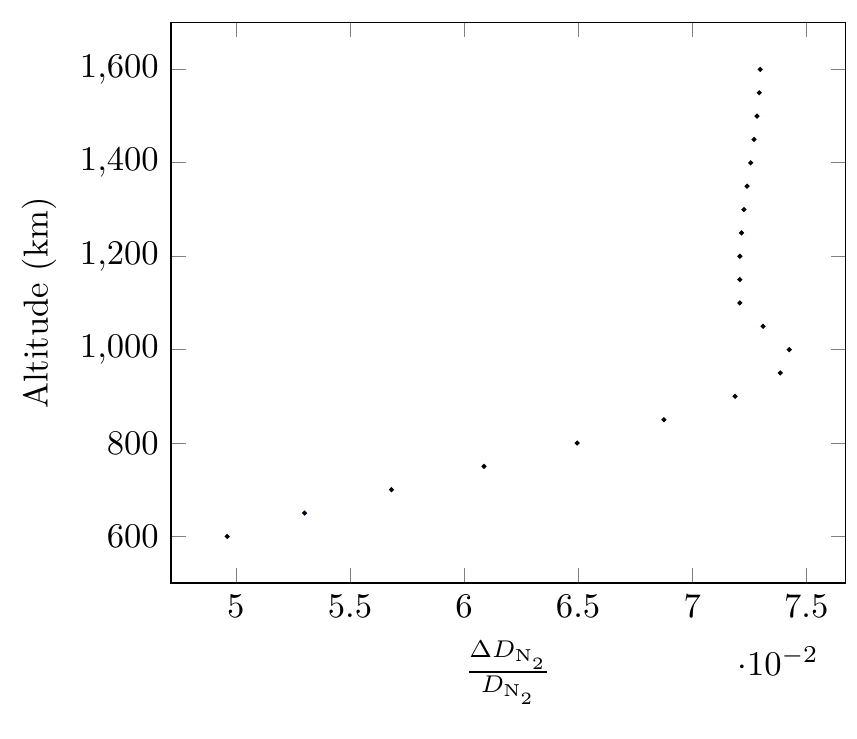}
\caption{\label{prog-unc_D}Wilke diffusivity relative uncertainty of \ce{N2}.}
\end{minipage}
\end{figure}

\begin{figure}
\centering
\includegraphics{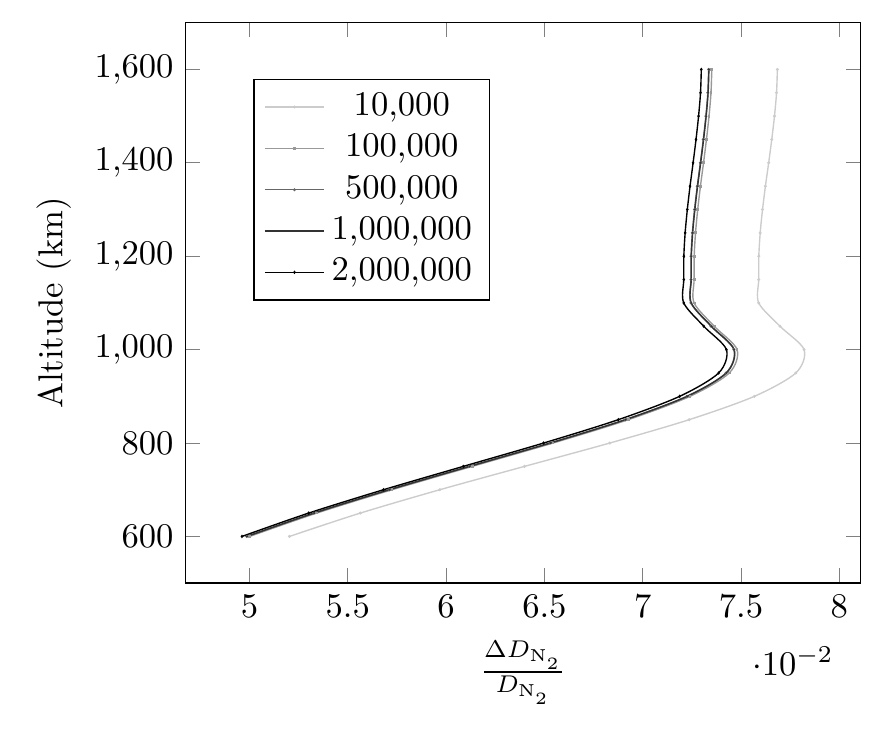}
\caption{\label{rel_conv}Convergence of the relative
uncertainty of the molecular diffusion coefficient of \ce{N2}.}
\end{figure}

The propagated values of \DCN\
are given in Fig.~\ref{prog_D12}, and its relative uncertainty is shown
in Fig.~\ref{prog-unc_D12}.
The mixture diffusion coefficient of \ce{N2} and its
uncertainty are given in Fig.~\ref{prog_D} and Fig.~\ref{prog-unc_D}.
Fig.~\ref{rel_conv} shows the convergence of the relative
uncertainty of the molecular diffusion coefficient of \ce{N2}, which
was found to be the most difficult to converge, requiring as many
as $2\times10^6$ draws.


\section{Discussion}
We have calibrated the Massman parameters for the bimolecular diffusion
coefficient for the \CNcouple\ pair using an uninformed prior.
The posterior distribution is given by the product of the likelihood and the
prior (Eq.~\ref{diffusion:Bayes}). Therefore the posterior is informed
by data through the likelihood and previously known knowledge through
the prior distribution. The choice of the prior ensures a maximal
effect of the information of the data in the Shannon entropy sense.

As figures~\ref{post_sample} and \ref{posteriors} show,
the previous fits from \cite{Wakeham1973} and
\cite{Massman1998} are but one point in the posterior space.
The cross/dashed lines are \cite{Wakeham1973} values, 
the plus/solid lines are \cite{Massman1998} values.
Although the fits accurately
reproduce the diffusivity at Titan's temperatures, all information
about the uncertainty is lost.

We observe a negative correlation between the relative uncertainty of
the molecular diffusion of \ce{N2} and the temperature 
(Fig.~\ref{cond_T40} and \ref{prog-unc_D}). A sensitivity
analysis of the Massman model (Eq.~\ref{derivative}) yields a 
$\ln\left(T/\Tz\right)$ term with respect to the $\beta$
parameter that can explain this behavior.
Titan's temperatures being lower than the reference temperature of
the Massman model ($\Tz = 273.15$~K) means this log-term will diverge to $-\infty$,
therefore increasing the uncertainty of the bimolecular diffusion
coefficient (Fig.~\ref{prog-unc_D12}) with respect to $\beta$. A sensitivity of
the molecular diffusion coefficient with respect
to the $\beta$ parameter could thus explain the negative correlation
between the molecular diffusion relative uncertainty and the temperature.
\begin{equation}
\begin{split}
\frac{\partial \Dbin}{\partial \Dzo}  & = \frac{\Pz}{P}\left(\frac{T}{\Tz}\right)^\beta, \\
\frac{\partial \Dbin}{\partial \beta} & = \Dzo\frac{\Pz}{P}\left(\frac{T}{\Tz}\right)^\beta\ln\left(\frac{T}{\Tz}\right).
\end{split}
\label{derivative}
\end{equation}
One important consequence
is that at lower temperatures, investigations to reduce uncertainties
on molecular diffusion should focus on the $\beta$ parameter,
typically by characterizing its value at these temperatures.

\begin{figure}
\centering
\includegraphics{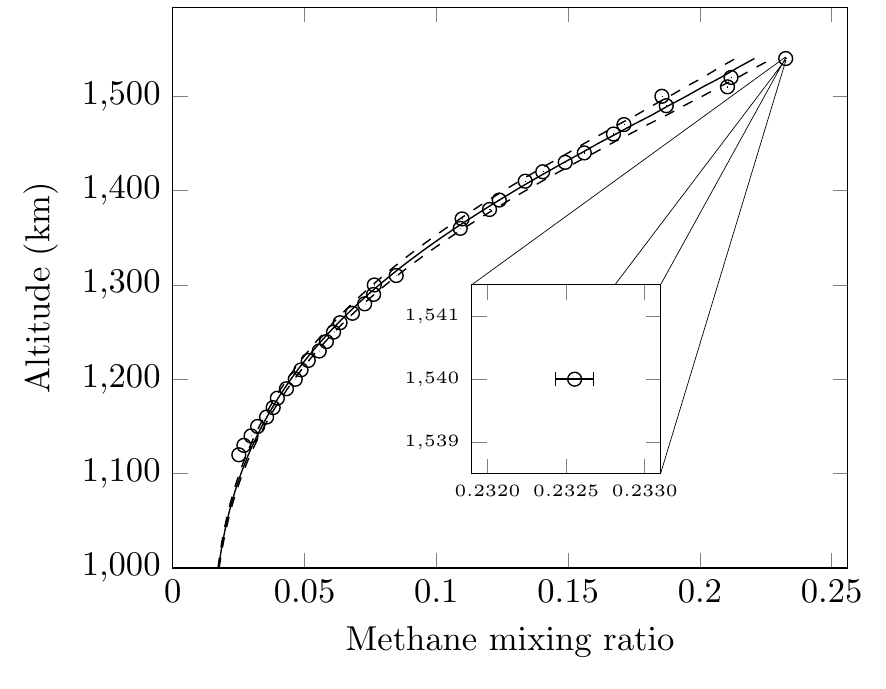}
\caption{Propagation of the uncertainties in molecular diffusion, roughly $\pm 7.5$\%,
to uncertainties in modeled methane mixing ratio. The data points are the measurements derived from CASSINI;
note that the statistical uncertainties from the INMS are orders of magnitude smaller.}
\label{meth_prop}
\end{figure}

In general, the uncertainties in molecular diffusion are small
compared to other sources of uncertainty, which can be orders of magnitude
larger \cite{Hebrard06, Carrasco07c}. However, a propagation of the uncertainties
in binary molecular diffusion to the calculation of the methane mixing ratio as
a function of altitude, illustrated in Fig.~\ref{meth_prop}, demonstrates that
the uncertainty is nontrivial above 1200~km.  This methane mixing ratio profile
is determined using an eddy coefficient that was optimized for the methane
profile and temperature and density profiles based on the T40 flyby as was done
in \cite{Mandt2012b}.  Also shown are the methane mixing ratios measured by
the Cassini Ion Neutral Mass Spectrometer \cite{Waite04} during the T40 flyby.
Uncertainties in molecular diffusion coefficients have the greatest impact at
the highest altitudes, because it is at these altitudes where molecular
diffusion has a dominant effect compared to eddy diffusion.  This has important
implications for studies of Titan's atmosphere that use models to evaluate
escape rates based on mixing ratios modeled above 1200~km \cite{Cui2012},
suggesting that uncertainties in molecular diffusion need to be taken into
account when making conclusions about measured altitude profiles.

For the common situation where no measurement data is available one has to
resort to other theoretical or model calculations.  In the case of missing
laboratory gas mixture data, diffusion coefficients may be estimated by a
transport theory calculation.  
The Bayesian framework can also handle this situation; the likelihood distribution will,
instead of acting on lab measurement data, act on transport theory calculation
output. 
It should thus capture the uncertainties and inadequacies associated to
the theoretical or model calculations. This requires a careful analysis 
of the corresponding model that is beyond the scope of this work.

In the situations where transport theory calculations yield poor diffusion
coefficient estimates, it is likely that the uncertainty in the diffusion
coefficients would be inflated.  Access to lab measurements offers the
possibility of providing favourable uncertainty calculations but investigating
the effect of theoretical calculations on diffusion coeffient uncertainty is a
topic of future work.

\section{Acknowledgements}
We acknowledge the financial help of NASA OPR
program with the grant number NNH12ZDA001N.

\bibliography{ref}
\bibliographystyle{plain}
\end{document}